\documentclass[sigconf]{acmart}

\copyrightyear{2026}
\acmYear{2026}
\setcopyright{cc}
\setcctype{by}
\acmConference[WiSec '26]{Proceedings of the 19th ACM Conference on Security and Privacy in Wireless and Mobile Networks}{June 30-July 03, 2026}{Saarbrücken, Germany}
\acmBooktitle{Proceedings of the 19th ACM Conference on Security and Privacy in Wireless and Mobile Networks (WiSec '26), June 30-July 03, 2026, Saarbrücken, Germany}
\acmPrice{}
\acmDOI{10.1145/3765613.3797453}
\acmISBN{979-8-4007-2201-1/2026/06}

\usepackage{svg}
\usepackage{enumitem}
\usepackage{longtable,booktabs}
\usepackage{array}
\IfFileExists{footnotehyper.sty}{\usepackage{footnotehyper}}{\usepackage{footnote}}
\makesavenoteenv{longtable}

\usepackage{multirow}
\usepackage{tablefootnote}
\usepackage{colortbl}
\usepackage{graphicx}
\usepackage{acro}
\usepackage{siunitx}
\usepackage{tablefootnote}

\usepackage{bbding}
\usepackage{fontawesome} 
\usepackage{transparent}
\usepackage{soul}
\setul{0.5ex}{0.3ex}  
\setulcolor{red}
\definecolor{lightgray}{gray}{0.925}

\DeclareAcronym{urllc}{
    short = URLLC,
    long  =  Ultra-Reliable Low Latency Communication
}
\DeclareAcronym{threegpp}{short=3GPP, long= 3rd Generation Partnership}

\DeclareAcronym{mmtc}{short= mMTC, long=Massive Machine-Type Communication}

\DeclareAcronym{sbi}{short= SBI, long=Service Based Interface}

\DeclareAcronym{ipsec}{short=IPsec, long= Internet Protocol Security}

\DeclareAcronym{up}{short =UP, long=User Plane}

\DeclareAcronym{cp}{short =CP, long=Control Plane}

\DeclareAcronym{ran}{short =RAN, long=Radio Access Network}

\DeclareAcronym{5gc}{short=5GC, long = 5G Core}

\DeclareAcronym{ue}{short=UE,long=User Equipment}

\DeclareAcronym{nf}{short=NF,long=Network Function}
\DeclareAcronym{gnb}{short=gNB, long=Next Generation Node B}

\DeclareAcronym{upf}{short=UPF,long=User Plane Function}

\DeclareAcronym{amf}{short=AMF,long=Authentication and Mobility Function}

\DeclareAcronym{smf}{short=SMF,long=Session Mangement Function}

\DeclareAcronym{pcf}{short=PCF, long = Policy Control Function}

\DeclareAcronym{lte}{short=LTE, long = Long Term Evolution}

\DeclareAcronym{tls}{short=TLS, long = Transport Layer Security}

\DeclareAcronym{ike}{short=IKEv2, long = Internet Key Exchange Version 2}

\DeclareAcronym{esp}{short=ESP,long = Encapsulation Security Payload }

\DeclareAcronym{psk}{short=PSK, long = Pre-Shared Keys}

\DeclareAcronym{nrf}{short=NRF, long = Network Repository Function}

\DeclareAcronym{supi}{short=SUPI, long= Subscription Permanent Identifier}

\DeclareAcronym{sim}{short =SIM, long = Subbscriber Identity Module}

\DeclareAcronym{qos}{short=QoS, long = Quality of Service}

\DeclareAcronym{bbu}{short=BBU, long = Baseband Unit}

\DeclareAcronym{rrh}{short=RRH, long= Remote Radio Head}

\DeclareAcronym{cu}{short=CU, long= Control Unit}

\DeclareAcronym{ru}{short=RU, long= Radio Unit}

\DeclareAcronym{du}{short=DU, long= Distributed Unit}

\DeclareAcronym{cuup}{short=CU-UP, long= Control Unit-User Plane}

\DeclareAcronym{cucp}{short=CU-CP, long= Control Unit-Control Plane}

\DeclareAcronym{nas}{short=NAS , long= Non-Access Stratum}

\DeclareAcronym{rrc}{short=RRC, long = Radio Resource Control}

\DeclareAcronym{cpri}{short=CPRI, long =  Common Public Radio Interface}

\DeclareAcronym{ecpri}{short=eCPRI, long =  enhanced CPRI}

\DeclareAcronym{mno}{short=MNO, long= Mobile Network Operator}

\DeclareAcronym{aead}{short=AEAD, long=Authenticated Encryption with Associated Data }

\begin{document}

\title{Evaluation of Security-Induced Latency on 5G RAN Interfaces and User Plane Communication}

\author{Sotiris Michaelides}
\email{michaelides@spice.rwth-aachen.de}
\orcid{0009-0003-6020-3934}
\affiliation{%
  \institution{RWTH Aachen University}
  \city{Aachen}
  \country{Germany}
}

\author{Jakub Lapawa}
\email{jakub.lapawa@rwth-aachen.de}
\orcid{0009-0000-8696-5492}
\affiliation{%
  \institution{RWTH Aachen University}
  \city{Aachen}
  \country{Germany}
}

\author{Daniel Eguiguren Chavez}
\email{daniel.eguiguren@rwth-aachen.de}
\orcid{0009-0004-6011-0372}
\affiliation{%
  \institution{RWTH Aachen University}
  \city{Aachen}
  \country{Germany}
}

\author{Martin Henze}
\email{henze@spice.rwth-aachen.de}
\orcid{0000-0001-8717-2523}
\affiliation{%
  \institution{RWTH Aachen University}
  \city{Aachen}
  \country{Germany}
}
\additionalaffiliation{%
  \institution{Fraunhofer FKIE}
  \city{Wachtberg}
  \country{Germany}  
}

\renewcommand{\shortauthors}{Michaelides et al.}

\begin{abstract}

  5G promises enhanced performance—not only in bandwidth and capacity, but also latency and security. Its ultra-reliable low-latency configuration targets round-trip times below 1\,ms, while optional security controls extend protection across all interfaces, making 5G attractive for mission-critical applications.
  A key enabler of low latency is the disaggregation of network components, including the RAN, allowing user-plane functions to be deployed nearer to end users. However, this split introduces additional interfaces, whose protection increases latency overhead.
  In this paper, guided by discussions with a network operator and a 5G manufacturer, we evaluate the latency overhead of enabling optional 5G security controls across internal RAN interfaces and the 5G user plane. To this end, we deploy the first testbed implementing a disaggregated RAN with standardized optional security mechanisms.
  Our results show that disaggregated RAN deployments retain a latency advantage over monolithic designs, even with security enabled. However, achieving sub-1\,ms round-trip times remains challenging, as cryptographic overhead alone can already exceed this target.

\end{abstract}



\maketitle

\section{Introduction}

The fifth generation of mobile networks (5G) introduces a major architectural transition, shifting from hardware-centric, monolithic deployments to software-defined, cloud-native designs~\cite{3gpp_ts_38_300_v18_7_0}. 
This evolution enables 3GPP, for the first time, to extend its focus beyond traditional consumer connectivity and to address industrial and mission-critical sectors that demand ultra-low latency~\cite{michaelides2025industry5G}.

In such sectors, the modular and cloud-native architecture of 5G—enabled by the disaggregation of the \ac{5gc} and \ac{ran} into individual network functions—allows latency-critical \ac{up} components (e.g., the \ac{cuup} and \ac{upf}) to be deployed closer to end devices, thereby reducing latency~\cite{michaelides2025industry5G,samsung2019vran,sajid2025towards} and enabling 5G to target Round-Trip Times (RTT) as low as \SI{1}{\milli\second}~\cite{latencyURLLC2025}.

Beyond low latency and high reliability, these critical sectors also demand strong security measures as their importance in societal, economic, and public safety-critical functions makes them a frequent target for powerful attacks~\cite{stellios2018survey, henze2013maintaining}.
Consequently, as 5G is being deployed in these environments---expanding the attack surface through new interfaces and components---strong security controls must be implemented to mitigate resulting threats.

Fortunately, 5G introduces a comprehensive suite of security controls across its architecture, covering nearly every component and interface. Although each component is required to support these mechanisms, the activation of most remains optional, granting network operators flexibility~\cite{michaelides2025industry5G,MichaelidesLatency,bicmac,tlscore}.
As enabling these controls inevitably introduces cryptographic operations that add latency, it remains open whether 5G can still achieve its promised sub-\SI{1}{\milli\second} RTT targets while maintaining robust security.

The importance of this issue is highlighted by the extensive prior work examining security controls within the \ac{5gc}, N3, and Uu interfaces~\cite{tlscore,MichaelidesLatency,bicmac}. 
While these studies provide insights into individual controls, they focus on isolated interfaces without considering the broader implications of \ac{ran} disaggregation, which is critical for low-latency communication~\cite{samsung2019vran,michaelides2025industry5G}. 
Thus, there is a need to evaluate security controls across the entire 5G architecture and their overall impact on system performance.

In this paper, we present the first comprehensive evaluation of optional security controls across the 5G \ac{up}, with a focus on disaggregated \ac{ran} deployments, crucial for achieving ultra-low latency~\cite{michaelides2025industry5G,samsung2019vran}. 
Our work is guided by discussions with a \ac{mno} and a 5G vendor, providing practical insights into the benefits of disaggregated \ac{ran} deployments and the use of optional security controls in real networks, ensuring that our findings have direct practical relevance.
More specifically, our key contributions are:
\begin{enumerate}[topsep=0pt]
  \item Practical insights from an \ac{mno} and a 5G vendor on security and performance in disaggregated \ac{ran} deployments, offering guidance for secure low-latency deployments~(\S\ref{sec:ranbenefits}).
  \item The first \emph{open-source} 5G testbed supporting all optional internal RAN security controls (\S\ref{sec:testbed}); available at~\cite{5GInternalRanSecurity}.
  \item Evaluation of the impact of security on latency across RAN interfaces for both control and user planes (\S\ref{sec:randeploy}), identifying IPsec as the most latency-efficient option, with configurations adding less than \SI{60}{\micro\second} of overhead.

  \item End-to-end \ac{up} latency analysis from \ac{ue} to \ac{upf} in monolithic and disaggregated setups~(\S\ref{sec:urllc}), showing that security alone already pushes RTTs beyond \SI{1}{\milli\second}.
\end{enumerate}

\begin{figure}[t] 
  \centering
  \includegraphics[width=0.475\textwidth]{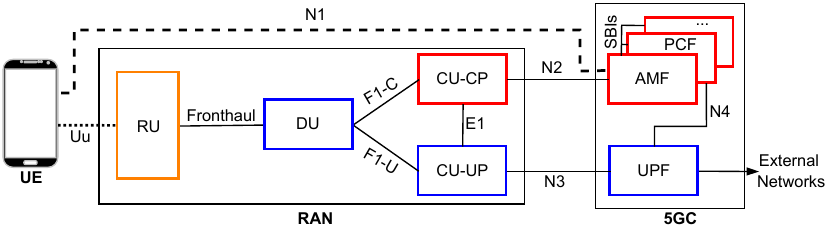} 
  \caption{The modularity of 5G enables time-critical components to be placed closer to end devices on the {\color[HTML]{0000FF}edge cloud} infrastructure to reduce latency. Control-plane components are hosted on a {\color[HTML]{FF0000}remote and centralized cloud} to simplify network control. The \ac{ru} is always deployed {\color[HTML]{FF8000}on-site}}
  \label{fig:5G}
\end{figure}

\section{5G Networks and Security Controls}
From a high-level view, 5G follows the same basic structure as earlier mobile networks, consisting of the \ac{ue}, the \ac{5gc}, and the \ac{ran} (cf. Fig.~\ref{fig:5G}).
It also maintains a split between the \ac{cp}, which handles signaling such as session management and mobility, and the \acf{up}, which handles user-data forwarding~\cite{MichaelidesLatency}.
This section outlines the 5G architecture (§\ref{sec:architecture}), the main interfaces and their associated security controls (§\ref{sec:security}), and the optional security mechanisms examined in related work~(§\ref{sec:related}).

\subsection{Architectural Components}
\label{sec:architecture}
The \ac{ue} is the end device that accesses network services provided by the \ac{5gc} using authentication credentials~\cite{ts23501}.  
The \ac{5gc} consists of interconnected \acp{nf} responsible for control and user-plane operations: \ac{cp} functions such as the \ac{amf} and \ac{smf} manage authentication, mobility, and session control, while the \ac{upf} on the \ac{up} handles routing of user data~\cite{ts23501}.  
The \ac{ran} mediates between the \ac{ue} and the \ac{5gc} by providing the wireless connectivity, and can be deployed either monolithically or disaggregated~\cite{ts38401}.  

In monolithic \ac{ran} deployments, the \ac{bbu} implements the full protocol stack except radio transmission, handled by the \ac{ru}. 
In a disaggregated deployment, the \ac{bbu} is split into \ac{cu} and \ac{du}~\cite{ts38401}. 
The \ac{du} handles real-time lower-layer functions, including scheduling, signal processing, modulation, coding, and beamforming, while the \ac{cu} manages higher-layer functions such as resource control, security, intra-\ac{ran} mobility, and \ac{du} management. 
The \ac{cu} is split into \ac{cucp} and \acf{cuup}, reflecting control- and user-plane separation, allowing independent scaling, optimization, and placement of each plane. 
Placing \ac{up} functions and the \ac{du} on the edge cloud ({\color[HTML]{0000FF} blue} in Fig.~\ref{fig:5G}) near the \ac{ue} is crucial for low-latency communication (cf.~§\ref{sec:ranbenefits}), while keeping \ac{cp} centralized ({\color[HTML]{FF0000} red} in Fig.~\ref{fig:5G}) enables better control.

\subsection{5G Interfaces and Security}
\label{sec:security}
Following the control–user plane separation, we examine the 5G interfaces and their associated security controls.

\textbf{Control Plane:}
\ac{cp} data is exchanged between network nodes to maintain and support 5G operations.
This includes signaling between the \ac{ue} and \ac{amf} over the N1 interface for authentication and mobility management, as well as signaling over the Uu interface between the \ac{ue} and the \ac{ran} for requesting and allocating radio resources.
These messages are mandatorily integrity-protected using the NIA algorithms based on SNOW, AES, or ZUC~\cite{bicmac, ts33501}, and may optionally be encrypted using the corresponding NEA schemes.
Within the \ac{ran}, control signaling occurs over the F1-C and E1 interfaces. 
These interfaces support operations such as intra-\ac{ran} handovers, the exchange of \ac{cp} data to and from the \acp{ue}, radio resource management, radio resource control forwarding, and \ac{up} control commands.
Control signaling between the \ac{ran} and the \ac{5gc} takes place over the N2 interface, which carries handover messages, mobility management information, and session-related control data between the \ac{cucp} and the \ac{amf}.
All three interfaces are optionally protected using IPsec or DTLS~\cite[§9.8.2, §9.8.3, §9.2]{ts33501}.
Within the \ac{5gc}, \ac{cp} data is exchanged between \acp{nf} for tasks such as session management and retrieving authentication credentials. 
Data over the \acp{sbi} can be optionally secured with TLS or IPsec~\cite[§13.1.0]{ts33501}, and with IPsec over the N4 interface~\cite[§9.9]{ts33501}.
While most security controls are optional \emph{"because they can be replaced with physical security"}~\cite{MichaelidesLatency}, disaggregated \ac{ran} deployments—where physical security is infeasible—often rely on these controls (cf.~§\ref{sec:ranbenefits}). 
Given that \ac{cp} security is critical for the network operation, we evaluate how security controls within the \ac{ran} affect crucial \ac{cp} operations of the 5G system~(§\ref{sec:randeploy}).

\textbf{User Plane:}
The \ac{up} is responsible for carrying the user data from the \ac{ue} to the \ac{upf}. 
Between the \ac{ue} and the \ac{ran} (\ac{cuup}), this data is optionally secured using the NIA/NEA schemes~\cite[§5.2]{ts33501}. 
While these controls also cover data over the F1-U interface between the \ac{cuup} and \ac{du}, security on this interface is crucial, as F1-U not only transports \ac{up} data but also carries control traffic used by the \ac{du} to regulate downlink data and prevent buffer overflows~\cite[§5.3]{ts38470}. 
IPsec is defined as an optional security control for this interface. The same applies to the N3 interface, which carries \ac{up} data between the \ac{cuup} and the \ac{upf}~\cite[§9.3]{ts33501}.
Security over \ac{up} interfaces adds latency and may affect 5G’s low-latency capabilities. 
In this paper, we first assess the impact of F1-U security in isolation~(§\ref{sec:randeploy}), as it is the only \ac{up} interface not studied in prior work~(cf.\,§\ref{sec:related}), before evaluating all \ac{up} security controls collectively~(§\ref{sec:urllc}).

\textbf{The special case of the Fronthaul (FH) Interface:}
The FH carries \ac{cp} and \ac{up} data and allows the \ac{du} to manage \acp{ru} (e.g., synchronization).
Defined by the \ac{cpri} consortium rather than 3GPP, proprietary CPRI links up to LTE transmitted analog signals, with security only feasible at the hardware level.
In 5G, \ac{ecpri} gradually replaces CPRI with a packet-based interface that enables partial distribution of \ac{du} functions to the \ac{ru}; however, implementing security controls (e.g., MACsec) is only recommended, not mandated, by the specification.
As the FH lacks standardized security controls and has diverse vendor-specific implementations, we exclude it from our evaluation, though we reference related work.

\begin{table}[t]
  \caption{While prior work examined the impact of various controls on isolated interfaces, our work is the first to evaluate the impact of security on (1) \colorbox{red!15}{RAN interfaces} and (2) \textcolor{blue}{UP interfaces} collectively to assess end-to-end latency overhead.}
  \label{tab:related}
  \centering
  \begin{tabular}{|c|c|c|c|c|c|c|c|}
  \hline
  \multicolumn{1}{|c|}{\textbf{-}} & \multicolumn{1}{c|}{\textbf{Interf.}} & \multicolumn{1}{c|}{\textbf{Security}} & \multicolumn{1}{c|}{\textbf{\cite{macsec}}} & \multicolumn{1}{c|}{\textbf{\cite{bicmac}}} & \multicolumn{1}{c|}{\textbf{\cite{MichaelidesLatency}}} & \multicolumn{1}{c|}{\textbf{\cite{tlscore}}} & \multicolumn{1}{c|}{\textbf{This}} \\ \hline
  
  \multirow{11}{*}{\rotatebox[origin=c]{90}{Control Plane}} 
  & \multirow{2}{*}{N1\&Uu} & NEA$^\beta$ & - & - & - & - & - \\ \cline{3-8}
  &  & NIA$^\alpha$ & - & - & - & - & - \\ \cline{2-8}

  & \multirow{2}{*}{SBIs} & TLS$^\beta$ & - & - & \checkmark & \checkmark & - \\ \cline{3-8}
  &  & IPsec$^\beta$ & - & - & \checkmark & - & - \\ \cline{2-8}
  
  & \multirow{2}{*}{N2} & DTLS$^\beta$ & - & - & - & - & - \\ \cline{3-8}
  &  & IPsec$^\beta$ & - & - & - & - & - \\ \cline{2-8}
  
  & N4 & IPsec$^\beta$ & - & - & - & - & - \\ \cline{2-8}
  
  & \cellcolor{lightgray} & \cellcolor{lightgray} DTLS$^\beta$ & \cellcolor{lightgray}- & \cellcolor{lightgray}- & \cellcolor{lightgray}- &\cellcolor{lightgray} - & \cellcolor{lightgray}\checkmark \\ \cline{3-8}
  &  \cellcolor{lightgray} \multirow{-2}{*}{\cellcolor{lightgray}\colorbox{red!15}{E1}} & \cellcolor{lightgray}IPsec$^\beta$ & \cellcolor{lightgray}- & \cellcolor{lightgray}- & \cellcolor{lightgray}- & \cellcolor{lightgray}- & \cellcolor{lightgray}\checkmark \\ \cline{2-8}
  
  & \cellcolor{lightgray} & \cellcolor{lightgray}IPsec$^\beta$ &\cellcolor{lightgray} - &\cellcolor{lightgray} - & \cellcolor{lightgray}- & \cellcolor{lightgray}- & \cellcolor{lightgray}\checkmark \\ \cline{3-8}
  & \multirow{-2}{*}{\cellcolor{lightgray}\colorbox{red!15}{F1-C}} & \cellcolor{lightgray}DTLS$^\beta$ &\cellcolor{lightgray} - & \cellcolor{lightgray}- & \cellcolor{lightgray}- & \cellcolor{lightgray}- & \cellcolor{lightgray}\checkmark \\ \hline
  
  \multirow{4}{*}{\rotatebox[origin=c]{90}{User Plane}} 
  & \cellcolor{lightgray}\colorbox{red!15}{\textcolor{blue}{F1-U}} & \cellcolor{lightgray}IPsec$^\beta$ & \cellcolor{lightgray}- & \cellcolor{lightgray}- &\cellcolor{lightgray}- & \cellcolor{lightgray}- & \cellcolor{lightgray}\checkmark + \\ \cline{2-8}
  
  & \cellcolor{lightgray}  & \cellcolor{lightgray}NEA$^\beta$ &\cellcolor{lightgray} - & \cellcolor{lightgray}- & \cellcolor{lightgray}- & \cellcolor{lightgray}- & \cellcolor{lightgray}+ \\ \cline{3-8}
  & \multirow{-2}{*}{\cellcolor{lightgray}\textcolor{blue}{Uu}} &\cellcolor{lightgray} NIA$^\beta$ & \cellcolor{lightgray}- &\cellcolor{lightgray} \checkmark & \cellcolor{lightgray}- & \cellcolor{lightgray}- & \cellcolor{lightgray}+ \\ \cline{2-8}
  
  & \cellcolor{lightgray}\textcolor{blue}{N3} & \cellcolor{lightgray}IPsec$^\beta$ & \cellcolor{lightgray}- &\cellcolor{lightgray} - &\cellcolor{lightgray} \checkmark & \cellcolor{lightgray}- & \cellcolor{lightgray}+ \\ \hline
  
  \multirow{2}{*}{\rotatebox[origin=c]{90}{Both}} 
  & \multirow{2}{*}{FH} & MACsec $^\gamma$ & \checkmark & - & - & - & - \\ \cline{3-8}
  &  & IPsec $^\gamma$ & - & - & - & - & - \\ \hline
  \end{tabular}
  $^\alpha$Mandatory Usage \hfill 
  $^\beta$Optional Usage \hfill 
  $^\gamma$Recommended Support \\
  + Combined Evaluation \hspace{2em}  \checkmark Isolated Evaluation
  \end{table}

\subsection{Related Work}
\label{sec:related}
The studies summarized in Tab.~\ref{tab:related} highlight the current state of research on optional security controls in 5G systems and motivate our work. 
Heijligenberg et al.~\cite{bicmac} were the first to examine the impact of optional security controls, evaluating NIA schemes on the \ac{up} in terms of bandwidth and latency and showing a noticeable yet comparable impact across all schemes. 
Zeidler et al.~\cite{tlscore} analyzed TLS as the primary security mechanism over the \acp{sbi}, reporting minimal performance degradation in an operational 5G system but significant delays during initial security establishment. 
Similarly, in prior work~\cite{MichaelidesLatency}, we investigated IPsec within the \ac{5gc} and across the N3 interface, demonstrating that IPsec can outperform TLS due to its ability to pre-establish security associations, while introducing only minimal latency on N3. 
Lastly, Dik et al.~\cite{macsec} examined MACsec (although not standardized) on the FH, showing a negligible impact on latency. 
Overall, these studies reveal the absence of an architecture-wide evaluation of the \ac{up} and a lack of consideration for disaggregated \ac{ran} deployments—vital for achieving low latencies—thereby underscoring the gap filled by our contributions.

\section{Insights on Disaggregated RANs}
\label{sec:ranbenefits}
To guide our research on disaggregated \ac{ran} architectures and optional security controls in real-world 5G deployments, we consulted a major 5G equipment manufacturer and a European \ac{mno} (who requested to not be named).
These discussions highlight the practical benefits of such deployments (§\ref{sec:mno}) and the use of optional security controls (§\ref{sec:manifact}), underscoring the relevance of this work.

\subsection{Practical Benefits of Disaggregated RANs}
\label{sec:mno}
During our discussions with the \ac{mno}, we learned that most current \acp{ran} deployments are monolithic due to legacy infrastructure. 
However, they emphasized that new \ac{ran} deployments are disaggregated and virtualized to improve cost efficiency, availability, and scalability. 
As they explained, it is more convenient to dynamically deploy instances of the \ac{du} or \ac{cu} rather than the entire stack, significantly reducing costs. 
These deployments also allow the \ac{mno} to quickly identify and isolate faulty instances and redeploy replacements, minimizing downtime—a top operational priority~\cite{MichaelidesLatency}.
Finally, the \ac{mno}, which is currently in the process of deploying a low-latency network slice, pointed out that internal measurements indicate that without physically reducing the distance between \ac{up} components and end-users, achieving low-latencies is infeasible.

\subsection{Security in Real-World 5G Networks}
\label{sec:manifact}
The equipment manufacturer we contacted is also responsible for configuring networks according to the requirements of clients, which include both MNOs and private companies.
The manufacturer reported an increasing demand for disaggregated and virtualized \acp{ran}, which aligns with findings from Dell’Oro (and Nokia) \cite{nokiacloudRAN}, projecting that by 2028, ~20\% of \acp{ran} will be disaggregated and virtualized.
When asked about security, the manufacturer noted that optional security controls are typically not used in traditional 5G deployments due to bandwidth considerations and potential performance bottlenecks especially on the \ac{up}, which bears more traffic.
In contrast, optional security mechanisms are almost always employed in disaggregated \ac{ran} deployments, as these components are often geographically distributed across cloud infrastructure where physical security is not feasible (cf.~§\ref{sec:security}).

Our discussions highlight the benefits of disaggregated \ac{ran} deployments, their role in low latency, and the use of optional security controls in real-world 5G networks, demonstrating the relevance and real-world applicability of evaluating security-induced latency on 5G RAN interfaces and \ac{up} communication.

\section{Deploying a Disaggregated-RAN 5G Testbed}
\label{sec:testbed}

To lay the foundation for such evaluations, we first identify optional security controls applied to the RAN interfaces (§\ref{sec:rancontrols}), before deploying them by extending an existing state-of-the-art testbed~(§\ref{sec:setup}).

\subsection{Identified Optional Security Controls}
\label{sec:rancontrols}
To identify optional security controls, we examine the relevant 3GPP specifications and determine the mandatory-to-support configurations, as these ensure a degree of compatibility between devices from different manufacturers.
For IPsec, we leverage the same controls as those we discovered in our previous work on evaluating tunneling protocols over the N3 and the SBIs~\cite{MichaelidesLatency}. 
These comprise six different algorithm combinations, which can be used with either certificates or pre-shared keys.
For DTLS, we extract three configurations from~\cite[§6.2]{ts33210}. 
All three configurations employ the same \ac{aead} algorithm for encryption and integrity protection (AES-GCM with 128-bit keys) but differ in their key exchange and authentication mechanisms. 
Interestingly, while standard TLS configurations mandate support for version 1.3, DTLS only requires version 1.2.

\begin{figure}[t] 
  \centering
  \includegraphics[width=0.475\textwidth]{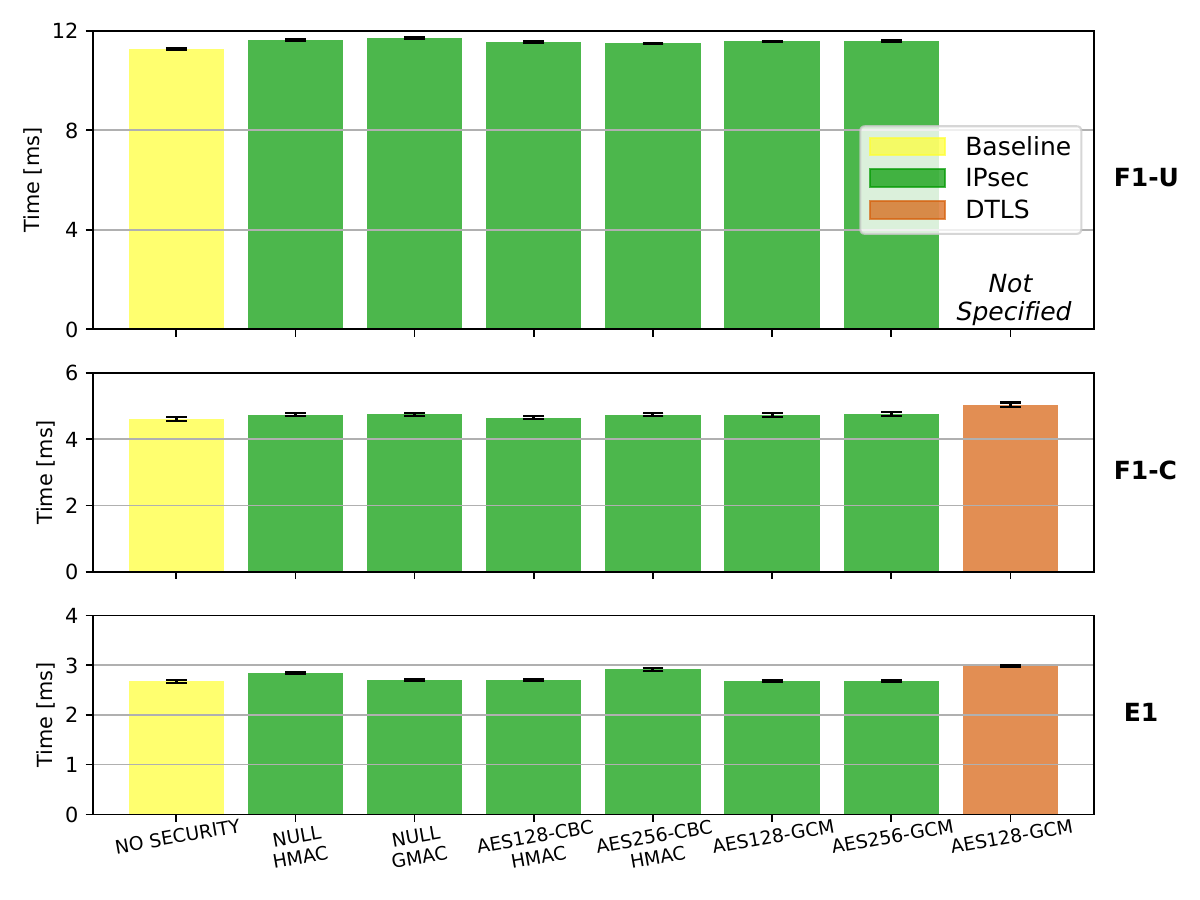} 
  \caption{Securing internal-RAN interfaces using IPsec adds only minimal latency overhead regardless of the configuration, making it suitable to support low-latencies.}

  \label{fig:Ran-Results}
\end{figure}

During our experiments, the \ac{ran} components establish their security (i.e., authentication and key exchange) once, during the discovery phase at deployment.  
Thus, our measurements focus on the overhead of encryption and integrity protection, evaluating six IPsec ESP configurations and one DTLS configuration (cf.~Fig.~\ref{fig:Ran-Results}).  
This mirrors real-world 5G systems, where the network runs continuously and security associations are thus rarely re-established.

\subsection{Testbed and Experimental Setup}
\label{sec:setup}

To deploy and study the identified security protocols, we extend our previously presented open-source testbed ~\cite{MichaelidesLatency}, which implements IPsec over the N3 interface using UERANSIM and Open5GS.
As UERANSIM lacks \ac{ran} disaggregation support and a full 5G protocol stack, we replace it with Open Air Interface (OAI)~\cite{oairan}, a widely adopted in academia and industry.
This upgrade not only supports a fully disaggregated \ac{ran}, but also allows seamless integration with software-defined radios, enabling both simulated and over-the-air deployments.
Each network function—including the \ac{5gc}, disaggregated \ac{ran} functions, and the \ac{ue}—runs as an independent Docker container, interconnected via dedicated subnetworks that emulate 3GPP interfaces.
The F1-C, F1-U, and E1 links are represented as separate Docker bridges, allowing their precise monitoring.

For consistency with the original setup, we deploy IPsec over the F1-C, F1-U, and E1 interfaces using strongSwan, establishing IKEv2/ESP tunnels automatically at container startup.
DTLS is implemented over the F1-C and E1 interfaces using the lightweight socat utility, which provides bidirectional DTLS tunnels between RAN components.
Our setup also allows dynamic switching between security configurations and authentication methods via environment variables, making it easily adaptable.
The complete testbed runs on a physical host with Ubuntu~22.04.5~LTS on an AMD Ryzen 7 Pro 6850U (8 cores, 16 threads) with an integrated Rembrandt GPU, 32~GB DDR5 RAM, and Linux kernel~6.8.0-60-generic.
The CPU supports AES-NI for AES acceleration, which we leverage to reflect realistic performance in modern 5G deployments.

\section{Impact of Security on Internal-RAN Interfaces}
\label{sec:randeploy}

To assess the impact of security on \ac{ran} interfaces, we use our testbed to perform reproducible \ac{cp} and \ac{up} tests, with and without security to quantify the overhead.  
We first outline the methodology we follow for each plane (\S\ref{sec:ranmeth}) before presenting our results~(\S\ref{sec:ranresults}).

\subsection{Methodology}
\label{sec:ranmeth}
 We use different methodologies for the two planes as we can directly generate traffic on the \ac{up}, whereas we cannot do so on the \ac{cp}.

\textbf{User Plane:} To assess the latency impact on the \ac{up} over the F1-U interface, we generate user data.
We employ the \emph{ping} tool, which uses ICMP messages to measure the RTT between two hosts, following the methodology of related work~\cite{bicmac,MichaelidesLatency}.
From within the \ac{ue} container, we ping the \ac{upf} container, ensuring that the traffic remains confined within the same physical host to minimize external factors that may affect latency.
We employ 1024-byte payloads to amplify the effect of cryptographic operations and repeat the experiment 20,000 times for statistical confidence.

\textbf{Control Plane:} In contrast to the \ac{up}, we cannot directly generate traffic on the \ac{cp} interface, as it is triggered by control procedures such as authentication and handovers.  
Therefore, we use \emph{UE registration}, one of the most common \ac{cp} operations~\cite{tlscore}, to trigger \ac{cp} activity.  
To this end, we deploy a \ac{ue} that automatically sends a registration request.  
While we do not monitor the registration data itself, we capture the \ac{cp} sub-procedures triggered during authentication to establish connectivity between the \ac{ue} and the \ac{upf}.  
Over the F1-C, we monitor the \ac{ue} Context Setup procedure, which establishes the \ac{ue} context and provides the \ac{du} with the information needed to manage the \ac{ue}.  
We measure the time from the \ac{ue} Context Setup request sent by the \ac{cucp} until the corresponding response, which indicates that the \ac{du} is ready to handle the \ac{ue}.  
For the E1 interface, we monitor the E1 Bearer Context Setup procedure, responsible for allocating resources and establishing GTP tunnels between the \ac{ue} and the \ac{upf} over \ac{cuup}.  
We measure the time between the Bearer Context Setup request from the \ac{cucp} to the \ac{cuup} and the corresponding response, confirming the successful establishment of GTP-U tunnels for the F1-U and N3 interfaces.  
To ensure statistical confidence, we repeat the experiment 1,000 times.

\subsection{Results} 
\label{sec:ranresults}

Our results (cf. Fig.~\ref{fig:Ran-Results}) are shown as 99\% confidence intervals of the average latency across all repetitions, providing statistical confidence in the true mean.  
Across all interfaces, the fastest IPsec configuration adds only minimal overhead on the order of tenths of microseconds, whereas DTLS consistently incurs higher latency.

\textbf{User Plane:}
Over the F1-U interface, where only IPsec is specified, all IPsec configurations combined add on average $\sim$\SI{120}{\micro\second} RTT ($\sim$\SI{60}{\micro\second} one-way), with minimal variation across cipher suites.  
Differences between cipher suites are indistinguishable, due to overlapping confidence intervals.  
These results closely align with related work on the \ac{up} (over N3), where the same IPsec algorithms were shown to add on average \SI{55}{\micro\second}, also with overlapping confidence intervals~\cite{MichaelidesLatency}.  
This indicates that IPsec is a strong candidate for secure, low-latency \ac{up} data transmission across the 5G architecture, as the added latency is well below the \SI{1}{\milli\second} threshold.

\textbf{Control Plane:}
On the \ac{cp}, specifically the F1-C interface, all IPsec configurations add on average $\sim$\SI{120}{\micro\second} to UE context setup, compared to $\sim$\SI{430}{\micro\second} for DTLS.  
The fastest IPsec configuration (AES256-HMAC) adds only $\sim$\SI{40}{\micro\second}, making it nearly indistinguishable from unprotected traffic.  
Over the E1 interface, IPsec configurations introduce on average $\sim$\SI{80}{\micro\second} of overhead, whereas DTLS adds $\sim$\SI{300}{\micro\second}.  
However, IPsec with AES128-GCM performs best, adding only $\sim$\SI{10}{\micro\second}.  
From a practical perspective, both protocols can be used for procedures that are not latency-critical—such as UE authentication—without noticeable impact.  
For latency-critical operations, such as intra-\ac{ran} handovers, IPsec remains the preferred option due to its consistently lower overhead.

\begin{figure}[t] 
  \centering
  \includegraphics[width=0.475\textwidth]{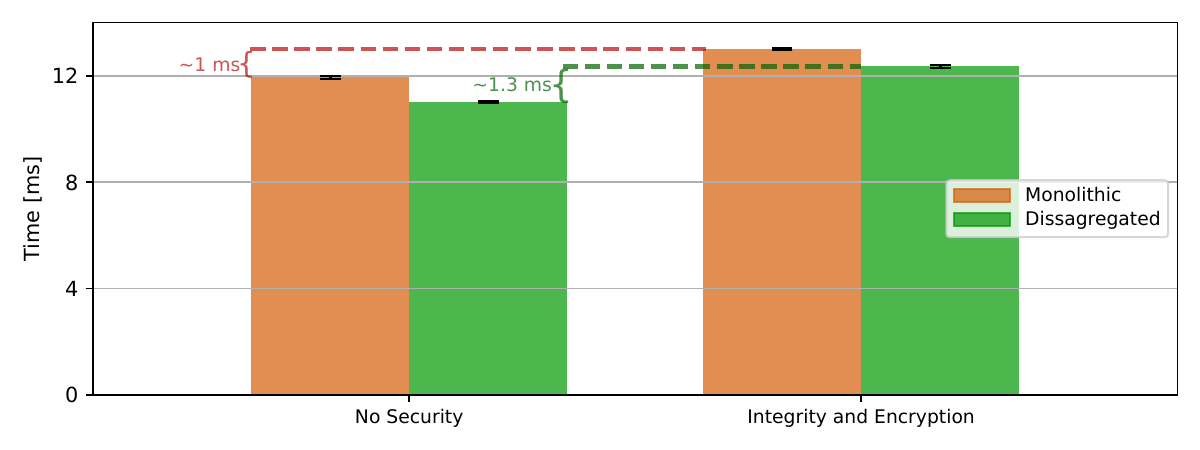} 
  \caption{While the monolithic deployment benefits from lower security overhead due to fewer controls needing to be enabled, the disaggregated approach still outperforms it due to the independent and efficient handling of \ac{up} data}

  \label{fig:Arch-wide-eval}
\end{figure}

\section{User Plane-Wide Evaluation}
\label{sec:urllc}
Having deployed and evaluated security within the \ac{ran}, we now assess its overall impact on latency when enabled across the entire 5G \ac{up}, from the \ac{ue} to the \ac{upf}.  
We first describe our methodology (\S\ref{sec:endmeth}), present results with 99\% confidence intervals (\S\ref{sec:rendresults}), and report tests conducted in a realistic 5G deployment (\S\ref{sec:realUe}).

\subsection{Methodology}
\label{sec:endmeth}
To evaluate the impact across the 5G \ac{up}, we follow a similar approach as for the F1-U measurements. We perform 20,000 ping measurements between the \ac{ue} and the \ac{upf} using 1024-byte packets.  
We assess latency overhead by comparing RTTs of a baseline without security to a deployment with security enabled on all \ac{up} links.  
We consider both a monolithic \ac{ran}, securing Uu and N3, and a disaggregated \ac{ran}, additionally securing F1-U.

To secure the interfaces, we have two options: either enable only integrity protection or enable both integrity and encryption. Encryption must always be integrity-protected for some interfaces (e.g., N3~\cite{MichaelidesLatency}) and, by itself, provides no protection against tampering attacks~\cite{breakingLTE}. 
However, we observed that enabling only integrity protection provides no measurable advantage over enabling both. 
Therefore, we choose to enable both encryption and integrity protection on all interfaces for enhanced security.
Over N3 and F1-U, we use IPsec with AES-GCM128, which provides both integrity and encryption, with performance comparable to schemes that only provide integrity protection. 
The same applies to the Uu interface in which we measure no difference when  enabling NIA and NEA simultaneously compared to using only NIA.
There, we utilize NIA\&NEA option 2 based on AES, as it has been shown to outperform the other two schemes in runtime and throughput~\cite{bicmac}.

\subsection{Results}
\label{sec:rendresults}
By examining the 99\% confidence intervals in Fig.~\ref{fig:Arch-wide-eval}, we observe that disaggregated \ac{ran} deployments achieve better baseline performance than monolithic ones (\SI{11}{\milli\second} vs.\ \SI{12}{\milli\second}), due to the independent handling of \ac{up} data—even without being physically closer to the end device. 
This confirms that disaggregated architectures are inherently well-suited for low-latency communication.
Even with security enabled, disaggregated \ac{ran} maintains its performance advantage despite the additional interfaces to protect. 
Cryptographic overhead exceeds \SI{1}{\milli\second} in both cases—approximately \SI{1}{\milli\second} for the monolithic setup and \SI{1.3}{\milli\second} for the disaggregated deployment. 
Given minimal CPU load and comparable memory usage, we attribute the measured overhead primarily to cryptographic operations.  
While disaggregated \ac{ran} remains faster overall, the high cost of cryptographic operations shows that achieving sub-\SI{1}{\milli\second} latency is extremely challenging for both \ac{ran} deployments, highlighting the combined effect of multiple optional security controls as a key limiting factor for ultra-low-latency applications.

\subsection{Validation in a Realistic 5G Deployment}
\label{sec:realUe}
To ensure reproducibility in an isolated setup, our main results are obtained using a simulated radio and \ac{ue}, where both disaggregated and monolithic \ac{ran} deployments are equally close to the \ac{ue}, minimizing the benefits of disaggregation.  
To validate these results in realistic 5G deployments, we repeat experiments using a \ac{ue} with a MediaTek Dimensity 700 baseband, a sysmocom ISIM-SJA5 SIM card, and a USRP B210, considering two scenarios differing in \ac{up} placement.  
In the disaggregated deployment, the \ac{upf} and \ac{cuup} are colocated with the \ac{du} on one host, placing \ac{up} functions closer to the \ac{ue}, while the \ac{cp} resides on a separate host connected via a network switch.  
In the centralized deployment, the \ac{du} remains local, while the \ac{up} components are colocated with the \ac{cp} on a second host, introducing ``distance'' from the \ac{ue}, resembling a monolithic setup.  
The disaggregated scenario achieves \SI{2.5}{\milli\second} and \SI{2.1}{\milli\second} lower latency without and with security enabled, respectively, highlighting the benefits of disaggregation for low-latency communication and validating the trends observed in our main experiments.

\section{IPsec Configuration Benchmarking}
\label{sec:benchmark}

While we focus on the latency of security controls, throughput is of equal concern to \acp{mno}, especially on the \ac{up} (cf.~§\ref{sec:manifact}).
Related work has studied NIA schemes, showing linear scaling with input size and identifying NIA2 as most efficient in throughput and runtime~\cite{bicmac}.
We complement these studies by benchmarking the impact of IPsec configurations using the Python Cryptography library.

\begin{figure}[t] 
  \centering
  \includegraphics[width=0.475\textwidth]{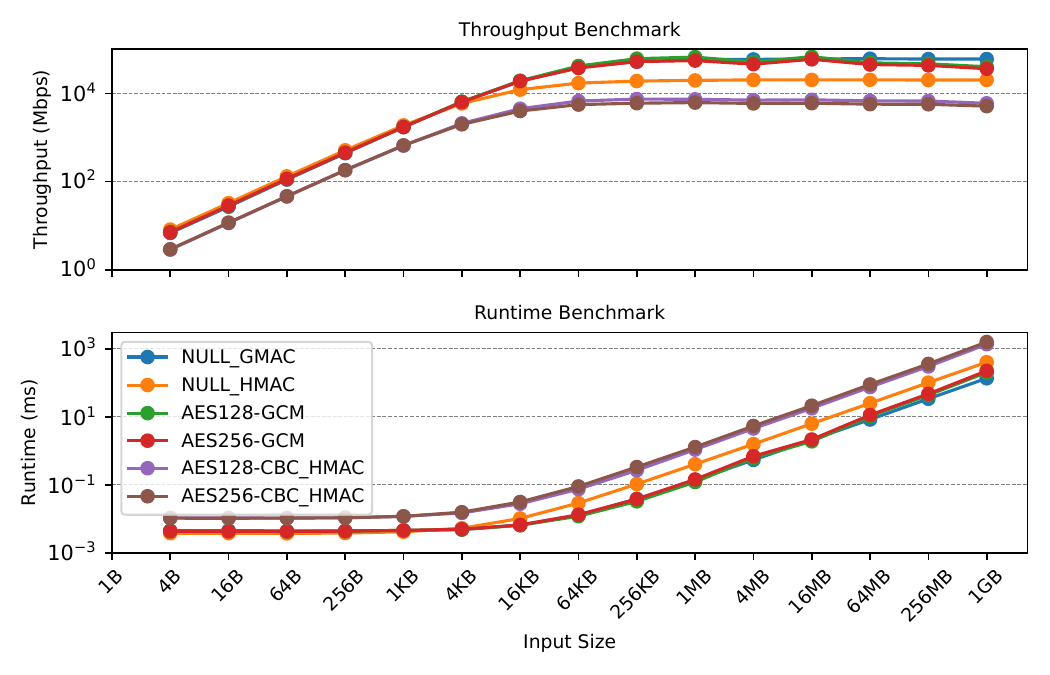} 
  \caption{Our benchmarking of IPsec configurations confirms the linear scalability of all schemes and highlights the superior performance of GMAC and GCM schemes.}
  \label{fig:bench}
\end{figure}

As shown in Fig.~\ref{fig:bench}, all ESP schemes scale linearly in throughput before capping at an upper bound.
GMAC achieves the highest throughput, reaching up to \SI{7.4}{\giga\byte\per\second} for large data.
The GCM schemes follow closely, averaging \SIrange{6}{6.5}{\giga\byte\per\second}, whereas CBC schemes remain the slowest, never exceeding \SI{1}{\giga\byte\per\second}.
These differences are reflected in runtime measurements: encrypting \SI{1}{\giga\byte} of data requires \SI{0.14}{\second} with GMAC AES-128, \SIrange{0.21}{0.23}{\second} with AES-GCM, and \SIrange{1.37}{1.61}{\second} with AES + HMAC, highlighting a \SIrange{6}{10}{\times} runtime advantage for GCM-based schemes.
The performance differences align with the underlying cryptographic constructions: 
GMAC is lightweight, AES-GCM combines encryption and GMAC in a parallelizable manner, while AES-HMAC is sequential and constrained by data dependencies. 
Consequently, operators can expect lower resource consumption and higher throughput when deploying GMAC or AES-GCM schemes.

\section{Discussion}
Finally, we discuss the broader implications of our results, including their transferability to real-world 5G systems, the underlying the root cause of high latencies, and potential avenues for improvement.

\noindent\textbf{Transferability of Results:}  
Even in our small-scale 5G network, latency already exceeds \SI{1}{\milli\second}, and overheads are likely to be higher in larger, real-world deployments with increased load. 
Our measurements show overheads in the order of tenths of microseconds—extremely fast—so further gains from more powerful machines would be minimal. 
Additional security controls, such as those over the FH or other non-parallelizable schemes (e.g., NIA1/NEA1 or AES-CBC\_HMAC), would further increase latency. 
Lastly, in our experiments, we ping the \ac{upf}, but in real 5G networks, where communication might occur between two \acp{ue}, cryptographic actions would double, effectively also doubling the overhead. 
As such, our conclusion that achieving sub-1 ms RTTs is challenging is broadly transferable to real-world 5G deployments.

\noindent\textbf{NIA/NEA as the Limiting Factor:}  
Although IPsec achieves very low latencies using fast, parallelizable schemes such as AES-GCM, the overall end-to-end latency remains high.
We attribute this high latency to the 5G NIA/NEA algorithms, which must be executed sequentially in a MAC-then-encrypt fashion \cite{bicmac}, inherently introducing additional latency.  
While NEA2 is based on AES-CTR, which is parallelizable, NIA2 relies on CMAC mode, which is  sequential, introducing data dependencies, further increasing latency.

\noindent\textbf{Reducing Latency:}
Latency can be reduced by minimizing cryptographic operations on \ac{up} data. Approaches include security-aware traffic separation (e.g., avoiding re-encryption of \ac{up} data over F1-U) or true end-to-end security between \ac{ue} and \ac{upf}. Lastly, using \ac{aead} algorithms (e.g., AES-GCM or ChaCha20‑Poly1305) in NIA/NEA schemes can further lower latency.

\section{Conclusion}
With this work, we underscore the importance of RAN disaggregation to meet low latencies in ultra-low-latency scenarios such as industrial networks.  
Guided by discussion with a real-world \ac{mno} and a \ac{ran} manufacturer, using a disaggregated \ac{ran} 5G setup, we show that the security impact on its interfaces can be kept to a minimum with IPsec, and---despite additional interfaces---it still outperforms its monolithic counterpart in an end-to-end \ac{up} evaluation.  
However, our results also reveal that achieving sub-millisecond RTTs remains difficult, as the cumulative cryptographic overhead already exceeds this threshold, even in a small-scale network.

\begin{acks}
  Funded by the German Federal Ministry of Research, Technology and Space (BMFTR) under funding reference number 16KIS2409K (6GEM+) and the German Federal Office for Information Security (BSI) under funding reference number 01MO24003B (CSII). The authors are responsible for the content of this publication.
\end{acks}

\bibliographystyle{ACM-Reference-Format-limit}

\end{document}